\documentclass[11pt]{article}

\usepackage{graphicx}
\usepackage{endfloat}
\usepackage[latin1]{inputenc}
\usepackage{amssymb,amsmath}
\usepackage{natbib}
\usepackage{color}
\usepackage{amsbsy}
\usepackage{bm}
\usepackage{titlesec}
\usepackage[english]{babel}

\newcommand{\Cov}{\textrm{Cov}}

\def\urltilda{\kern -.15em\lower .7ex\hbox{\~{}}\kern .04em}

\begin{document}
\bibliographystyle{jecol}

\baselineskip26pt
\phantom{MEE}

\vspace{3cm}
\begin{center}
{\huge \bf Dismantling the Mantel tests}\\
\bigskip
\bigskip
{\Large Gilles Guillot$^{1*}$  and Fran\c cois Rousset$^2$}\\
\end{center}

\vspace{3cm}

{\large
\noindent 1. Informatics and Mathematical Modelling Department,
Technical University of Denmark, Copenhagen.
Richard Petersens Plads, Bygning 305, 2800 Lyngby,  Denmark.\\
2. Institut des Sciences de
l'\'Evolution
(UM2-CNRS-IRD),
Universit\'e Montpellier 2,  Place Eug{\`e}ne Bataillon,  CC 065,  Montpellier
cedex 34095,  France.\\
Running title: Dismantling the Mantel tests.\\
\noindent $^*$ Corresponding author: gigu@imm.dtu.dk\\
Word count: 4000}

\newpage
\topmargin30pt
\centerline{\bf Summary}
{
\begin{enumerate}
\item The  simple and partial Mantel tests are routinely used in many areas of evolutionary biology to assess
the significance of the association
between two  or more matrices of distances   relative to the same pairs of individuals or demes.
Partial Mantel tests rather  than simple Mantel tests are widely used to assess the relationship between two variables displaying some form of structure.
\item We show that contrarily to a widely shared belief, partial Mantel tests are not valid in this case,
and their bias remains close to that of the simple Mantel test.
\item We confirm that strong biases are expected under a sampling design and spatial correlation parameter drawn from an actual study.
\item The Mantel tests should not be used in case auto-correlation is suspected  in both variables compared
under the null hypothesis.
We outline alternative strategies. The R code used for our computer simulations is distributed as supporting material.
\end{enumerate}
}

\bigskip
\bigskip

{\bf Keywords:}
 Association between two variables,  landscape ecology,  landscape genetics,  phylogenetics,
phylogeography,  geographic epidemiolgy, spatial structure,  isolation by distance, isolation by resistance, auto-correlation,
p-value,  type I error, {\em Loa loa},  onchocerciasis.

\newpage
\topmargin-45pt

\section{Introduction}

For  the detection of clustering of cancer cases in space and time,
\citet{Mantel67} introduced  a test based on permutations.
He concluded his article by claiming  that this method was general
- a claim later relayed  by \citet{Sokal79} - and could be used
whenever one has to assess the significance of the correlation between the entries of two square matrices
containing ``distances''  relative to pairs of individuals.
\citet{Dietz83} discussed the efficiency of various measures of correlation and
\citet{Smouse86} proposed an extension of the test,  referred  to as partial Mantel test,
and aimed at assessing the dependence between two matrices of distances while controlling
 the effect of a third distance matrix.
The latter may contain phylogenetic distances or plain geographical  (Euclidean) distances between pairs of sampling sites
but it may alternatively contain  values that attempt to reflect the actual cost for an individual to move
across the geographical area  (accounting e.g. for the presence of  barriers or hostile areas).
In the latter case,  the distance is known in ecology as ``cost distance''.
It  may not enjoy the properties of a mathematical distance (lacking the triangular inequality property)
but it is in general correlated with the Euclidean distance.

Since the original papers of \citet{Mantel67} and \citet{Sokal79},  and despite  the fact that (or perhaps because)
none of them stated the null hypothesis explicitly, the simple and partial Mantel tests have  known a tremendous popularity.
The Mantel tests are for example used routinely to assess the significance of the association
between two  matrices  of phenotypic or genetic distances.
They are also extensively used to assess  how a  matrix of genetic or phenotypic distances
relates to a matrix of geographical distances  \citep[see e.g.][and references therein]{Legendre10},
or to test if such distances can be explained by  phylogenetic relatedness.
Another classical analysis consists in assessing the significance of the dependence between genetic (or phenotypic) distances
and cost distances while ``controlling for the effect'' of geographical distances through the partial Mantel test
\citep[see e.g.][]{Cushman06, Storfer07, Balkenhol09b}.
In view of the various tasks above,  the Mantel tests have a number of appealing features.
First they allow one to synthesise information contained in multivariate data in a single index
and hence in a single test; second they allow one to deal with the case outlined above where the ``distance''
between individuals cannot be
expressed as a difference (or combination of differences) between one or several variables (e.g. case of a cost distance);
finally,  they do not seem to rely on any parametric assumption.

It is well known that data displaying some form of structure or auto-correlation are ubiquitous in ecology.
Their analysis brings up some statistical issues \citep{DinizFilho03}
 because structure or auto-correlation violates many assumptions made by
standard statistical methods.

Regarding the Mantel tests,  some concerns regarding the  type I error rate and power have been raised.
\citet{Oden92} reported a problem for the partial Mantel test for spatially auto-correlated data.
\citet{Lapointe95} found problems for the simple Mantel test when used for the comparison of dendrograms.
 \citet{Raufaste01} and \citet{Rousset02} gave an example  where the partial Mantel test leads to the wrong conclusion.
Lastly,  \citet{Nunn06} and \citet{Harmon10} expressed concerns about the simple and partial Mantel tests
when used for phylogenetic comparative analyses.
Essentially, all the issues reported by these studies relate to inflated type I error rate or low power.

In a recent review article, \citet{Legendre10} discussed some of these issues but on
the basis of simulations concluded that the simple and partial Mantel tests were valid statistical
methods without any recommendation about the conditions of validity.
In another study,  \citet{Cushman10} carried out a simulation study of performances
of the simple and partial Mantel tests in a set up typical of landscape genetics studies.
Under the simulation condition considered, they claimed that, in contrast to the simple Mantel test,
the type I error rates of the partial Mantel test was not inflated. However,
they only reported average correlation coefficients, not actual error rates of the tests.

There is therefore a great confusion about these tests and despite the various criticisms expressed about them,
they are still routinely used in many branches of evolutionary biology.
In the present article we
(i) clarify what assumptions are involved in the use of the simple and partial Mantel tests,
(ii) investigate by simulation the effect on the  tests  of structure (auto-correlation) in the data and show that
under some widely encountered conditions and a broad range of parameter values,
the simple and partial Mantel tests do not achieve the targeted type I error rate,
(iii) analyse theoretically the source of the problem, in particular emphasising that it results
from the permutation procedure common to all variants of partial Mantel tests  
(iv) outline how existing  methods for structured data could be used when the Mantel tests are not valid.

\newpage
 \section{Material and Method}

\subsection{The Simple and Partial Mantel Tests}

The procedure  introduced by \citet{Mantel67} is as follows:
for a data-set $(x_i, y_i)_{i=1, ..., n}$
(i) compute the set of pairwise distances $D^x_{ij}=\mbox{dist}(x_i, x_j)$  and $D^y_{ij}=\mbox{dist}(y_i, y_j)$;
(ii) compute the scalar product $r= \sum D^x_{ij}D^y_{ij}$;
(iii) for a large number of times:
(iii-a) draw a random permutation of $\{1, ..., n\}$ uniformly,
(iii-b) compute the  set of pairwise distances $\tilde{D}^x_{ij}$ for the vector of permuted $x_i$s,
(iii-c) compute the scalar product  $\tilde{r}= \sum \tilde{D}^x_{ij}D^y_{ij}$;
(iv) derive a p-value by comparing $r$ to the collection of $\tilde{r}$ values obtained in (iii).\\
The partial Mantel test introduced by \citet{Smouse86} aims at controlling for the effect of a third distance matrix $D^S$.
In this method,  the test statistic $r$ is the partial correlation coefficient
of $D^x$ and $D^y$ given $D^S$,  the permutation procedure remaining the same.

\subsection{Modelling Framework}

\subsubsection{A model for one variable structured in space}
In a first class of widely encountered problems,  $x_i$ is a  coordinate that locates individual or deme $i$
(hereafter referred  to as unit $i$ for short) in the geographical space
or on a phylogenetic tree and $y_i$ is a set of genetic or phenotypic observations relative to  unit $i$.
In this case,  $x$ can be treated  as a deterministic variable and $y$ as a (random) function of the variable $x$.
Testing the dependence of $y$ on $x$ can be done by testing whether $y(x)$ is a random function that displays some form of auto-correlation.
This is not the only way to model the dependence between two variables but it is one way which is relevant to many studies in
evolutionary biology.

\subsubsection{A model for two variables structured in space}
In a second set of problems,  both $x_i$ and $y_i$ are phenotypic or genetic observations about unit $i$.
In this case,  there is no reason to give a different status (random versus deterministic) to $x$ and $y$ as they
play the same role in the analysis.
In this second setting,  it is more natural to model both $x$ and $y$ as random functions.
These random functions depend on a third variable  $s$  which is deterministic,
again typically,  a coordinate locating an individual on the geographical space or on a phylogenetic tree.
In this case,  testing the dependence between $x$ and $y$ amounts to testing the dependence between
two random functions that are both potentially auto-correlated.

A key  point of the present paper is that under the two-random-function model and in presence of auto-correlation,
both the simple and partial Mantel tests fail to return the targeted type I error rate.
This will be illustrated by a simulation study and then analysed from a more theoretical point of view.

\subsubsection{An explicit model to simulate under $H_0$}
\paragraph{A rich and parsimonious family of random functions: the Gaussian random field model}
The present simulation study is concerned with evaluating the type I error rate of the Mantel tests
for two auto-correlated random functions.
Defining $H_0$ as ``$X$ and $Y$ are independent'' is not enough.
To be able to analyse the statistical properties of the tests,  we need a statistical model
allowing us to compare   $r$ to the distribution of $\tilde{r}$.
We need a model that is simple enough to be simulated and analysed easily but rich enough
to encompass the main features encountered in real life.
A widely used model for auto-correlated variables  whose variation is observed throughout the geographical space is the Gaussian random field model (GRF).
A Gaussian random field $x(s)$ is a function of the geographic coordinate $s$ such that
for any set of locations $(s_1, ..., s_n)$,  ${\bf x} = (x(s_1), ..., x(s_n))$ forms a multivariate Gaussian random vectors,  i.e.
has a probability density function  (pdf) proportional to
$\displaystyle \exp[-\frac{1}{2}({\bf x} - {\bm \mu})^t\Sigma^{-1}({\bf x} - {\bm \mu})]$
\citep{Mardia79}.
In informal terms: ${\bf x}$ has the well-known bell-shaped pdf centred around ${\bm \mu}$,  that can be easily visualised in one or two dimensions.

\paragraph{Simplifying the Gaussian random field model:}
The variation of a  random field can be described by its mean function $\mu(s)$ and its covariance function
$\Cov[x(s), x(s')]$ between arbitrary locations $s$ and $s'$.  
The mean function represents the expected value at geographical site $s$. If the random field models a variable which can be
observed repeatedly (e.g. annual cumulative rainfall at site $s$), $\mu(s)$  can be thought of as the average value
across multiple observations at site $s$.
If the random field represents a variable that is essentially unique (e.g. elevation) then $\mu(s)$
is introduced for modelling convenience and is often assumed to have a simple parametric form
(e.g. linear trend).
 For the sake of model parsimony,  we assume here that  $ \mu(s)$ is constant or linear in $s$.
The covariance function quantifies the linear statistical dependence between $x(s)$ and $x(s')$ as a function of $s$ and $s'$.
For $s$ and $s'$ far apart, it is reasonable to assume that $x(s)$ and $x(s')$ become independent.
The covariance function allows us to model at which rate and which distance this decorrelation occurs.
Again,  for the sake of model parsimony,  we assume
that  $\Cov[x(s), x(s')]$ depends only on the geographical lag  $s-s'$ between $s$ and $s'$
and not on the direction of $s-s'$.
This set of assumptions about $\mu(s)$ and $\Cov[x(s), x(s')]$ is known as second-order stationarity.
Further,  we assume that $\Cov[x(s), x(s')]$ depends only on the geographical distance $h = |s-s'|$
and not on the orientation of the vector $s-s'$.
This property is known as isotropy.
To summarise,  we have here a stationary,  isotropic Gaussian random field model.
This  is widely considered as a parsimonious,  yet flexible and powerful model to study variables structured in space
\citep{Chiles99, Lantuejoul02, Diggle03, Wackernagel03, Gelfand10}.
Among the broad family of parametric models of covariance functions currently used,  we chose the exponential model,  i.e.
we assume that the statistical dependence between $x(s)$ and $x(s')$,  as measured by the covariance,
decays exponentially as a function of the geographical distance
between $s$ and $s'$: $\Cov[x(s), x(s')] = \exp(-|s-s'|/\kappa)$.
In the above,  $\kappa$ is a scale parameter and has the dimension of a geographical distance.
We assume a similar model for $y$ and we assume that $x$ and $y$ are independent.

\subsubsection{Graphical examples of two independent stationary isotropic GRFs}
\begin{figure}[h]
\begin{tabular}{l} 
\vspace{0cm}\hspace{0cm} \includegraphics[width=8cm]{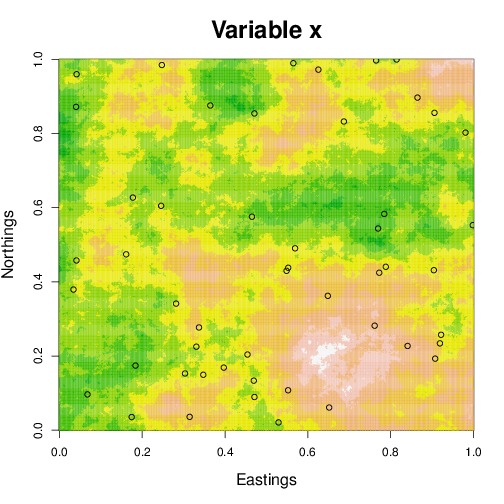}
\vspace{0cm}\hspace{0cm} \includegraphics[width=8cm]{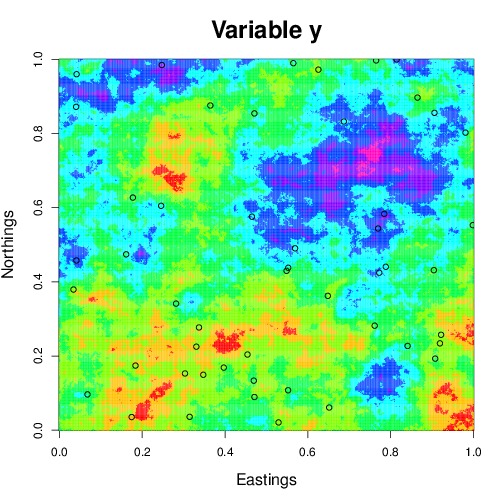} \\
\hspace{4cm}(a) \hspace{8cm}(b) \\
\vspace{0cm}\hspace{0cm} \includegraphics[width=5.2cm]{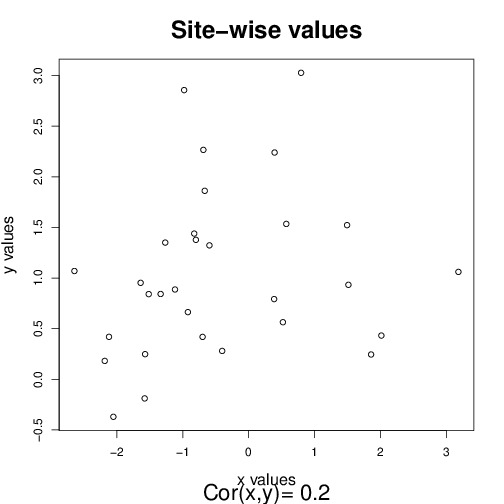}
\vspace{0cm}\hspace{0cm} \includegraphics[width=5.2cm]{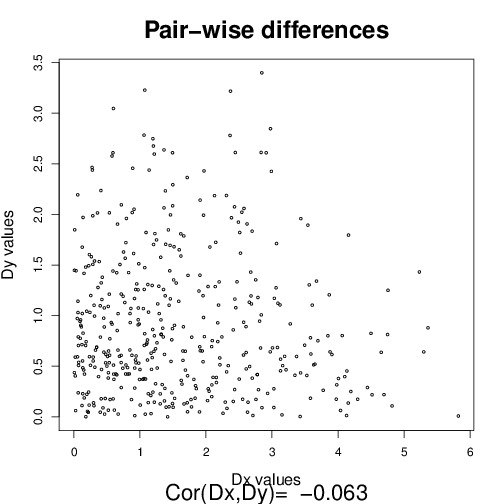}
\vspace{0cm}\hspace{0cm} \includegraphics[width=5.2cm]{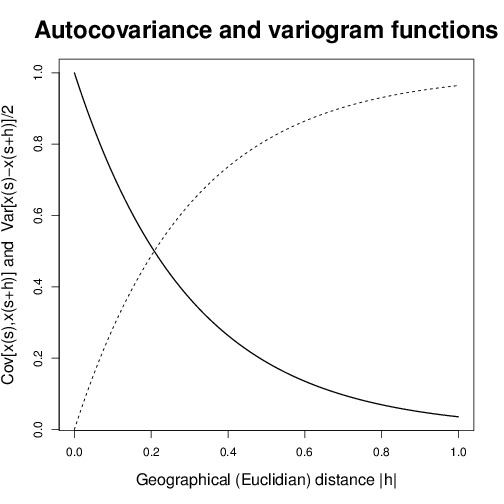} \\
\hspace{2.5cm} (c) \hspace{5cm}(d) \hspace{4.5cm} (e)
\end{tabular}
\caption{Top:  two independent random fields with an exponential covariance function with a common scale parameter $\kappa=0.3$
and the locations of 50 sampling sites; Because the covariance function decays slowly, values taken at pairs of close sites
are similar. For a spatial lag $s-s'$ large enough the value at a site $s$ does  not say much about that value at site $s'$,
there is statistical decorrelation.
Bottom:
 scatter plot of individuals $x$ values versus $y$ values at the 50 sampling sites(c),
scatter plot of pairwise distances $|x_i-x_j|$ versus $|y_i-y_j|$ values (d),
common theoretical covariance and variogram functions of $x$ and $y$ (e).}\label{fig:image}
\end{figure}

An example of realisations of our model on a fine grid
is shown on figure \ref{fig:image}.
For all pairs of sites $(s, s')$,  $\Cov[x(s), x(s')]$ and  $\Cov[y(s), y(s')]$ are non zero.
For this reason,  the variations of $x$ and $y$  both  display a clear structure in space.
Each variable could represent a genetic variable (e.g. the logit transform of an allele frequency for a population with continuous spread
in space),  a phenotypic variable or an environmental variable.
If $y(s)$ represents an environmental variable such as the elevation or the temperature at sites $s$,
then a matrix of pair-wise differences $D^y$ could be interpreted as a matrix of ecological cost distances.

\subsubsection{Simulation study design}
Biological or environmental variables are often not observed in the continuum but rather at a
limited number of  irregularly spaced sites. We considered here the case where $x$ and $y$ are observed
at the same set of $n=50$ sites or $n=200$ sites .
We considered the cases $\kappa=0,  \kappa=0.3$ and $\kappa=0.7$.
These values are meaningful only relative to the diameter of the sampling window.  
With $\kappa=0.3$, decorrelation is approximately reached  at one distance unit which is
the window width here.

For each value of $\kappa$,  we simulated 200 data sets
$(x_{1}, ..., x_{n})$,  $(y_{1}, ..., y_{n})$,  computed the Mantel statistics $r$ with 10000 permutations
and computed the associated ``p-value'' following Mantel's algorithm.
In case auto-correlation is suspected,  a recommended strategy consists in
 entering the matrix $D^s$ of geographical distances in a partial Mantel test between  $D^x$ and $D^y$
with the aim of ``controlling the effect of distances''.
We implemented this strategy with the same simulation design as above.
We also carried out the same experiment with random fields displaying a  linear spatial trend $\beta^t s$.
Such a trend could arise from the presence of large scale geographical features (e.g. distance to the sea) in
the spatial variation of temperature.
We sampled the two-component parameter $\beta$ of the linear trend from an independent two-dimensional normal  ${\cal N}(0, 1)$ distribution.
Because some authors have discussed the effect of various permutation procedures \citep{Smouse86, Legendre00} we implemented
four different permutations strategies,  referred to as Methods 1--4 in \citep{Legendre00} and in figures \ref{fig:qqplot_all}
and \ref{fig:qqplot_all_n_200}.
There procedures differ in the nature of the statistic computed and exchanged among pairs of locations, but are all instantiations of the same permutation procedure originally defined by Mantel.

\clearpage
\pagestyle{plain}
\section{Results}

\begin{figure}[h]
\begin{tabular}{c}
\vspace{-.3cm}\hspace{-.6cm} \includegraphics[width=17.cm]{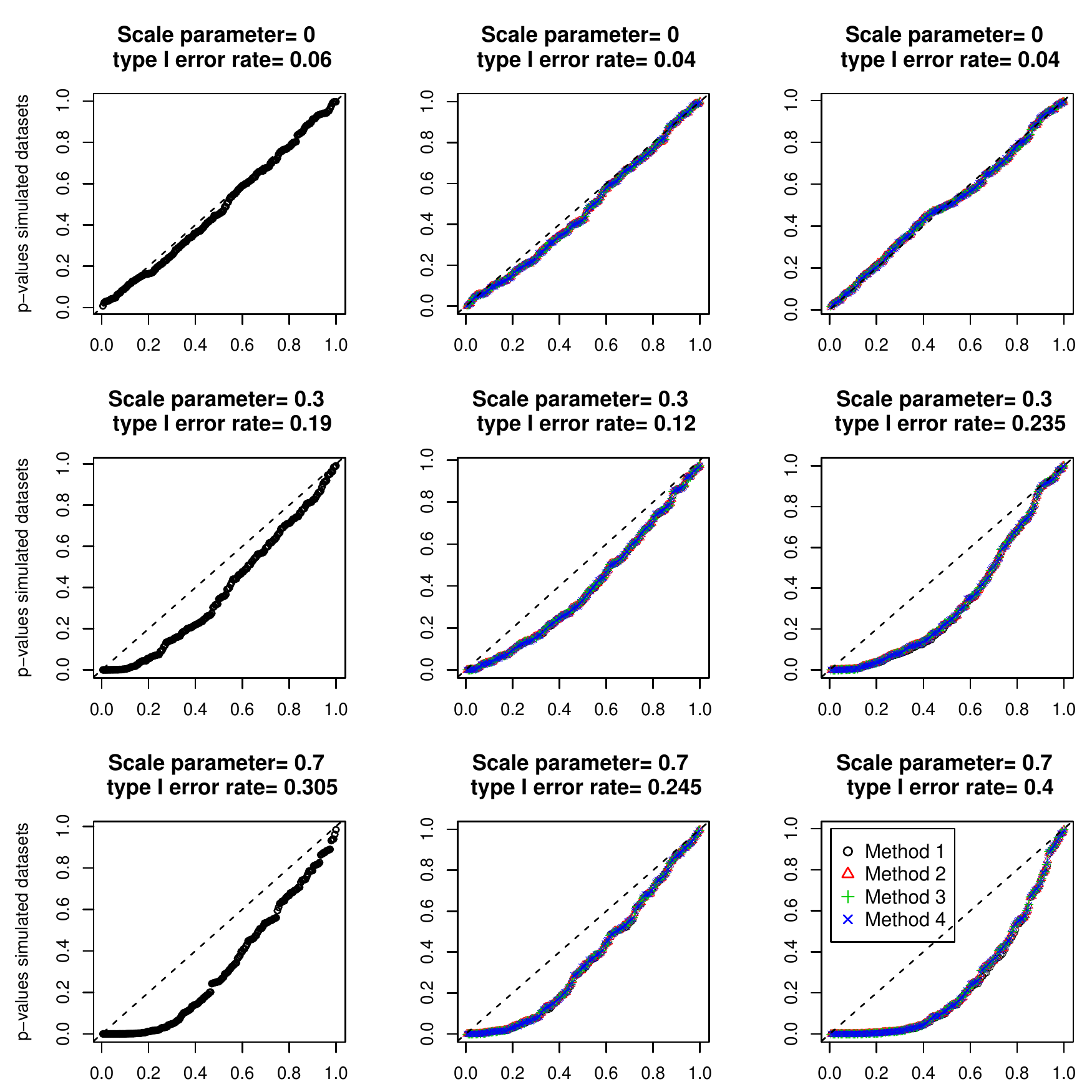} \\
\end{tabular}
\caption{Quantile-quantile plots of  p-values obtained on simulated data with $n=50$ sampling sites.
Each point corresponds to a simulated data set. The y-axis is the p-value returned by the simple Mantel test and the x-axis
is the corresponding quantile for a uniform distribution on $[0, 1]$.
The null hypothesis  tested is the independence between $x$ and $y$.
Left column: simple Mantel test,  the matrices $D^x$ and $D^y$ are obtained from independent random fields with zero mean (no deterministic spatial trend).
Middle column: partial Mantel test,  the matrices $D^x$ and $D^y$ are obtained from independent random fields with zero mean (no deterministic spatial trend).
Right column: partial Mantel test,   the matrices $D^x$ and $D^y$ are obtained from independent random fields with a deterministic linear spatial trend.
For the partial Mantel test (middle and right columns),  each data set was analysed by four permutation methods.
The p-values should be aligned along the diagonal.
The type I error rate reported is achieved for a targeted level of $\alpha=0.05$.
See also main text for references about methods 1-4.}
\label{fig:qqplot_all}
\end{figure}
\begin{figure}[h]
\begin{tabular}{c}
\vspace{-.3cm}\hspace{-.6cm} \includegraphics[width=17.cm]{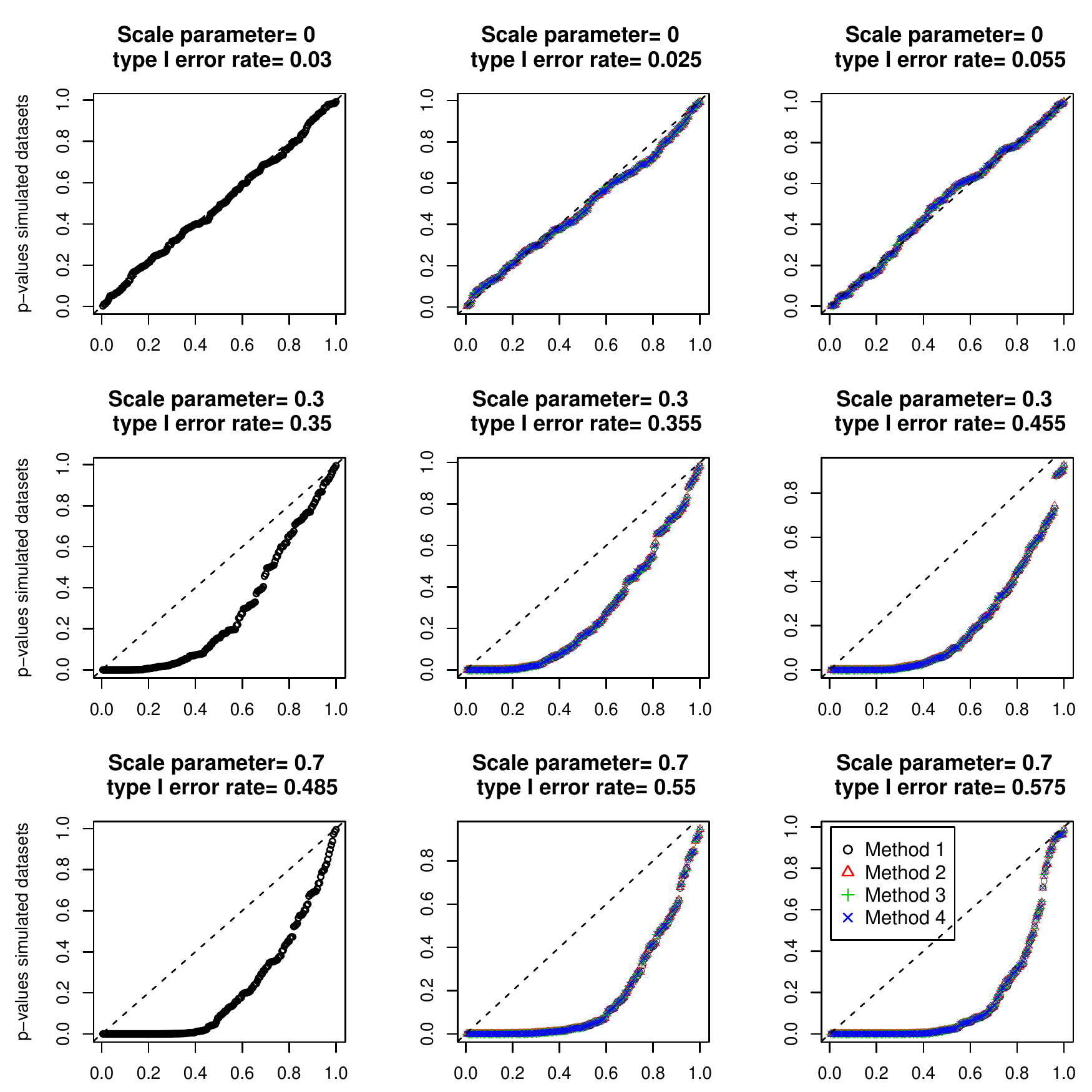} \\
\end{tabular}
\caption{Quantile-quantile plots of  p-values obtained on simulated data with $n=200$ sampling sites.
See Fig.~\ref{fig:qqplot_all} for details.}
\label{fig:qqplot_all_n_200}
\end{figure}

The proportion of false positives for a test at level $\alpha$ should be $\alpha$ for any $\alpha \in [0, 1]$.
In other words,  the   distribution of the p-values obtained should be uniformly distributed. 
The distribution of the p-values obtained are illustrated on figure \ref{fig:qqplot_all}.
Misalignment with the diagonal line shows a departure of the p-values with the expected uniform distribution
hence a problem in the test investigated.
For $\kappa=0$,  i.e. in  absence of spatial auto-correlation in both variables,  the simple and partial Mantel tests perform well (top row).
If spatial structure is present in the data in form of a deterministic linear trend only (no auto-correlation),
this can be controlled by introducing a third matrix of geographic distances in a  partial Mantel test
(cf Fig.~\ref{fig:qqplot_all} top right panel).
As soon as spatial structure is present in form of auto-correlation (spatial scale parameter $\neq 0$),
the simple Mantel test fails to achieve the targeted type I error rate.
It produces indeed  a considerable excess of small p-values i.e. the simple Mantel test rejects
the null hypothesis of independence too often and produces a much higher number of false positives than what it should do.
In the presence of auto-correlation,  including the matrix of geographic distances in a  partial Mantel test  
does not ``control'' auto-correlation as the  excess of small p-values remains substantial.
Besides,  the curves for the four methods
investigated by 
\citet{Legendre00} are perfectly overlapping (cf Fig.~\ref{fig:qqplot_all} middle and right columns).
This shows that the failure of the partial Mantel test is not inherent in the choice of one of the
particular variants of the permutation procedure discussed earlier in the literature.
The magnitude of the excess of small p-values increases with the magnitude of auto-correlation in the data and
for the highest spatial scale parameter value considered here ($\kappa=0.7$),  the type I error rate effectively achieved
for a test targeting $\alpha=5\%$ ranges between 25\% and 40\%.
It increases with the sample size, up to 55\%\ when $n=200$ (Fig.~\ref{fig:qqplot_all_n_200}).

\clearpage

\begin{figure}[h]
\begin{tabular}{c}
\vspace{-.3cm}\hspace{3cm} \includegraphics[width=10.cm]{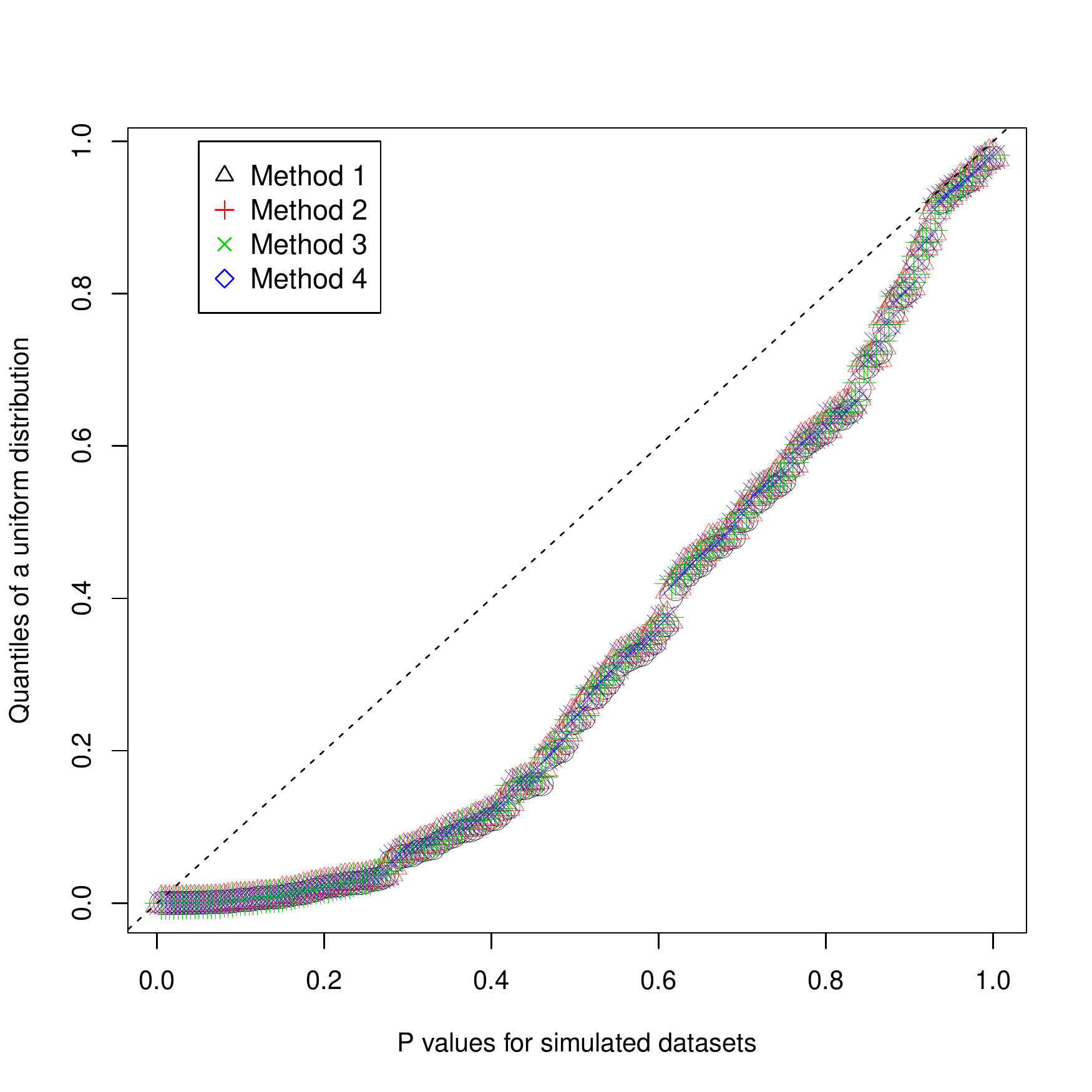} \\
\end{tabular}
\caption{Quantile-quantile plots of  p-values obtained when the $x$ variable is  the elevation observed
at $n=196$ sites in a 500km $\times$ 500km  area in Cameroon and the $y$ variable
is  simulated independently from $x$ but auto-correlated  with the same spatial characteristic scale
 as the prevalence of {\em Loa loa}
infection in this area \citep[as per parameter estimates  by][]{Diggle07b}.
The type I error rate is $27.5\%$\ at the nominal $\alpha=5\%$}
\label{fig:qqplot_loaloa}
\end{figure}

The realism of the design of the simulation study can be questioned. For example, the correlation among the most distant samples remains large when $\kappa=0.7$ in our simulations. However, large biases of the tests are also observed in realistic conditions. Few biological studies provide estimates of $\kappa$, and we here consider \citet{Diggle07b} who investigated the effect of elevation and a vegetation index on the prevalence of infection by the filarial nematode {\it Loa loa} (involved in onchocerciasis) in Cameroon. We have therefore assessed the performance of a partial Mantel test of the effect of elevation on a response variable having the same auto-corrrelation 
as reported in this work ($\kappa=0.7$ in units of longitude and latitude, but with more distant samples in these units than in our previous simulations), the same number (196) and positions of sampled locations (which were more clustered than in our previous simulations), and the same observed values of elevation. Out of 200 replicate simulations, the realized error rate was 27.5\%\ at the nominal 5\%\ rate (Fig.~\ref{fig:qqplot_loaloa}), showing that large biases can be observed in real conditions.

\section{Discussion}

\subsection{Why are the Mantel tests returning erroneous p-values?}

\subsubsection{Simple Mantel test}

Let us define $H_0$ as ``$x$ and $y$ are uncorrelated''.
The question is now: does the simple Mantel permutation procedure produce correlation coefficient
values according to the right distribution?
The distribution of the correlation coefficient involves not only the dependence structure between $x$ and $y$ but also
the joint distributions of $(x_{1}, ..., x_{n})$ and that of  $(y_{1}, ..., y_{n})$. So the answer depends
on what these joint distributions are.
If $(x_{1}, ..., x_{n})$ and $(y_{1}, ..., y_{n})$ are both independent and identically distributed (i.i.d),  then permuting the entries of $(x_{1}, ..., x_{n})$ breaks the potential
dependence between $x_i$ and $y_i$ while leaving the joint distribution of $(x_{1}, ..., x_{n})$ unchanged.
We note here for the sake of completeness that the i.i.d assumption above is not strictly required
and the assumption of exchangeability,  i.e. invariance of the distribution under permutation of its variables
\citep{Kingman78} is sufficient.

In the case of correlated data,  the consequences of the permutation are different.
For a small spatial lag $|s_i-s_j|$ and because of the spatial structure,
$x_{i}-x_{j}$ and $y_{i}-y_{j}$ tend to have the same order of magnitude (typically $x_{i}-x_{j}$ and $y_{i}-y_{j}$
are both small),  even though the random fields $x$ and $y$ are independent.

The effect of a permutation amounts to substitute index $k$ to index $j$ for various pairs $(j,k)$.
After permutation, the correlation coefficient computed involves the pair of sites $(i,j)$ in
a term which is actually $(x_{i}-x_{k})(y_{i}-y_{j})$ (up to some centring).
The term $(x_{i}-x_{k})$ has no reason to be of the same order of magnitude as $(y_{i}-y_{j})$.
In other words, the permutation not only breaks
the potential dependence between $x$ and $y$  but also breaks the spatial structure among the entries of the
variable subject to permutation.

The permutation procedure has no reason to produce  values from the distribution of the test statistics  under the null hypothesis.
The simple Mantel test produces values typical of the correlation coefficient between two independent variables
when one of them is not structured. What is required for a proper test is
 the distribution of the correlation coefficient between two independent variables
being both structured. 
In landscape genetics, one is often interested in testing whether a genetic data-set $x$ for a species
subject to isolation-by-distance (IBD) display variation that can be explained by environmental variable(s) $y$.
The Mantel test will produce pseudo genetic data by re-sampling. These pseudo-genetic data will be (as targeted) independent
of the environmental conditions $y$ but there will no longer be representative of IBD data.
There will be more likely to look like data arising from an island model.

\begin{figure}[h]
\begin{tabular}{c}
\vspace{-.3cm}\hspace{-1cm} \includegraphics[width=17.cm]{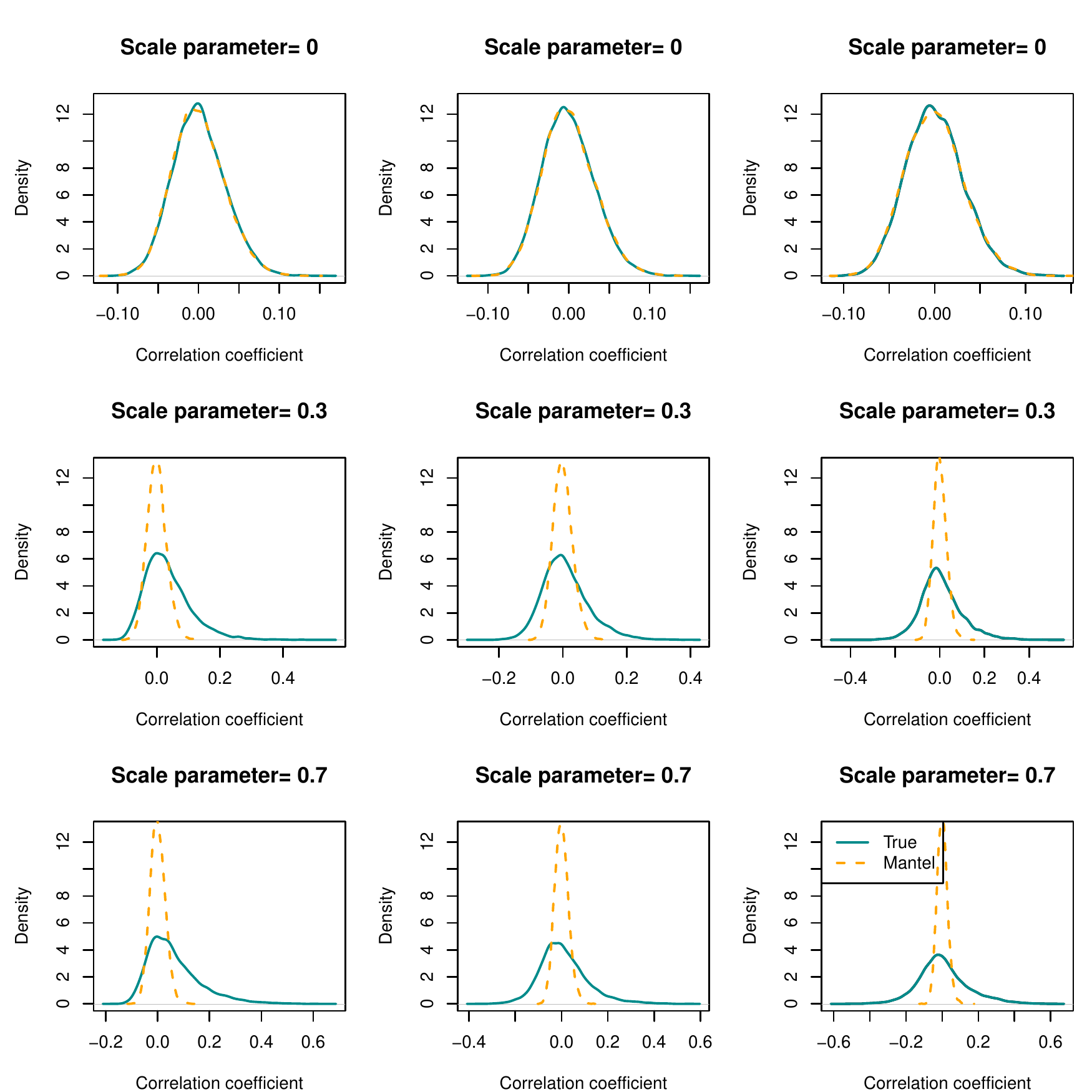} \\
\end{tabular}
\caption{Estimated density functions for the true distribution of the correlation coefficient
and the Mantel correlation coefficient between two variables
at  $n=200$ sampling sites as in section ``Simulation study design'' and Fig. \ref{fig:qqplot_all_n_200}.
In presence of auto-correlation, the Mantel  distribution is erroneous and
displays far less dispersion than the true distribution.
}
\label{fig:density}
\end{figure}

The simple Mantel permutation procedure produces values that display far less dispersion than what
{\em auto-correlated  data} should do  under the null hypothesis (as illustrated Fig.\ref{fig:density}).
In presence of auto-correlation,  the feature of the data  that is implicitly rejected is not the independence of $x$ and $y$
but rather the absence of spatial structure of $x$ and $y$.

\subsubsection{Partial Mantel test}
In the partial Mantel test,  the test statistic is the partial correlation coefficient of $D^x$ and $D^y$ given $D^s$.
This coefficient is defined as  the usual correlation coefficient between the residuals of $D^x$ and $D^y$
after a linear regression on $D^s$.
The variation of a random field is far more complex than a linear trend (cf. Fig. \ref{fig:image}-d).
The regressions of $D^x$  on  $D^s$ and of $D^y$ on  $D^s$ fail to capture most of the variability in the data.
The residuals of these regressions still display some spatial structure with an intensity that is close
to that present in the site-wise values.
The procedure consisting in permuting sampling units and computing correlation of residuals
is subject to the same issue as in the simple Mantel test.
The correlation coefficient values obtained by permutations do not display enough variability.

\subsection{When are the Mantel tests valid statistical methods?}

\subsubsection{Testing the existence of structure of one variable in space}
Let us consider the model outlined in section 'One random function' above and let us  assume that
the random function $y(x)$ is stationary (i.e. the statistical distribution of any finite sequence of $y(x_i)$ values
is invariant by spatial translation of the $x_i$ values).
Under this model,  the Mantel permutation procedure produces  correlation coefficient values from the distribution  under the null hypothesis.
The stationarity assumption above is a reasonable assumption in the case of genetic mutation-migration-drift equilibrium.
The simple Mantel test is therefore suitable to test the absence of isolation-by-distance from population genetic data in this case.

\subsubsection{Testing the dependence between two random variables}

When $x$ and $y$ are two random functions of a third variable,  the simple Mantel test is valid
if both $x$ and $y$ are  stationary and
either $x$ or $y$ is non-auto-correlated.
The latter situation has been partially studied by \citet{Legendre05} but we note that none
of the above assumptions is clearly relevant in evolutionary biology studies.

\subsubsection{Re-conciliating our results with findings of \citet{Legendre10}}

In their recent review, \citet{Legendre10} discussed at length the Mantel tests.
In section ``Multivariate, spatially structured data'' they  write:
"Permutation tests used in
the analysis of rectangular data tables by regression or
canonical analysis, or in the analysis of distance matrices
by Mantel tests, all have correct levels of type I error; so
they are all statistically valid."
Our results are in stark contrast with this claim. So what is going on?

This section above in \citet{Legendre10} reports findings from an earlier study by \citet{Legendre05}
summarised in Appendix 2 of Supplementary Material to \citet{Legendre10}.
They do not give enough detail about the simulation study but our analysis suggests unambiguously that either
(i) the model simulated by \citet{Legendre05} does not produce spatially auto-correlated data at all
or (ii) the parameters used (not clear from the original publication) 
are such that the data are structured
at a spatial scale that is much smaller than the sampling window.

Also, \citet{Legendre10} conclude their abstract by writing: ``The Mantel test should not be used as a general method for
the investigation of linear relationships or spatial structures in univariate or multivariate data.
Its use should be restricted to tests of hypotheses that can only be formulated in terms of
distances''.
Potential users are easily misled by such claims to consider that their problem  ``can only be formulated in terms of
distances'' and then feel allowed to use the (partial) Mantel tests.
Yet, it is not clear what these claims means. Indeed, it is proper to simulate the per-location data out of which matrices are computed, which means that hypotheses are de facto not expressed ``only in terms of distances'' but in terms of per-location data. This emphasizes that hypotheses are better described as features of probability distributions (e.g. a correlation function), and that distances matrices are only a way of representing data, not a way of specifying hypotheses. Thus, no set of conditions where partial Mantel tests are valid is identified by such claims.

\subsubsection{Re-conciliating our results with findings of  \citet{Cushman10}}
In their study, \citet{Cushman10} simulated genotypes from a model accounting explicitly for the landscape
or the presence of a barrier and concluded that
the Partial Mantel test could accurately identify the proper scenario and was not prone to any
false discovery as reported in the present study.

However, they only reported average ``Mantel'' correlation coefficients in 100 simulations, which is not sufficient to
make conclusions about the validity of the partial Mantel test. For example in the conditions where partial Mantel tests performed worst in Fig.~\ref{fig:qqplot_all},
the average partial correlation coefficient was 0.005 (see also Fig.~\ref{fig:density}). 
This emphasizes that the whole distribution of correlation coefficients, not only its mean, matters for the validity of the test.
We do not know how partial Mantel tests would actually perform in their simulations.
Their study includes three hundred Monte Carlo simulations but they are all
drawn from a single landscape resistance surface (identified by \citet{Cushman06}
as the most supported out of 110 alternative models for black bear {\em Ursus americanus}
gene flow in northern Idaho, USA).
Rather than rehabilitating the partial Mantel tests, such a study would rather suggest a method of investigating {\em ex-post}
how realistic Mantel p-values are.
However, this involves intensive simulations and a similar computational burden
should allow scientists to perform better inferences along the lines suggested below,
and estimate directly key ecological parameters explaining genetic variation, without resorting to the Mantel tests.

\subsection{Alternative strategies}

\subsubsection{Testing independence between two point processes}
We first note that the original problem of \citet{Mantel67} amounts to detecting the dependence
between the marks and locations of a marked point process. This problem is not chiefly relevant to evolutionary biology data
as information about the sampling effort is unfortunately often not available.
 However,  we note that this question  has been  addressed in a rigorous way by \citet{Schlather04}.

\subsubsection{Shift permutation}
We recall that if at least one of the variables to be compared is observed at the nodes
of a regular grid (as it may occur in landscape genetics and phylogeography with Geographic Information System data),
one can apply shift permutations of this variable \citep{Upton95}.

\subsubsection{Adapting the t-test}
Assessing the significance of the correlation between two random fields is a question that has been
considered by \citet{Clifford89, Richardson91} and \citet{Dutilleul93}
for  quantitative continuous variables
and \citet{Cerioli02} for categorical variables.
The methods proposed there can be readily used when the
data are available as site-wise uni-variate values.
They can be also adapted when the data are multivariate.
The case where data are pair-wise distance matrices that are not obtained as differences between
site-wise values requires further work.

\subsubsection{Fixing the Mantel test for non-linear trends}
The incentive for developping the partial Mantel test was the intuition that one variable can has a confounding effect
when analysing the dependence between two other variables.
The method fails to fulfil its promise because the method was not based on an explicit statistical model
and the implicit underlying model (a simple regression) is most often not appropriate.
However,  the idea of filtering out some (possibly non-linear) deterministic trend to retrieve some transformed
data (under which the Mantel test is appropriate) could be further explored.
This could lead to methods for data with  spatial structure in form of deterministic trend only.

\subsubsection{Inference and testing in a hierarchical  model}


Linear-mixed models with auto-corrrelated errors are correctly-specified models for analyzing data simulated
by Gaussian random fields, so that likelihood ratio tests based on such models should be at least asymptotically valid.
Analysing the dependence between two variables can be considered in the broader
framework of a generalised linear mixed model (GLMM), allowing non-Gaussian response variables.
Inference and testing in such  models can be made e.g. with Monte-Carlo methods \citep{Robert04},
or adjusted profile likelihood methods \citep{Lee09}.
Simulations based methods akin to Approximate Bayesian Computation \citep[e.g.,][]{Marin11}
could also be useful.
The framework of hierarchical models coupled with modern numerical inference and testing methods
is promising but it has not not been investigated as an alternative to the Mantel tests so far.
Although procedures for fitting spatial GLMM models have been described in the literature (e.g. \citealp{Diggle07b,Lee12}),
there is a dearth of computer implementations to performs such analyses in an automated way.
Most GLMM softwares, for example, do not include procedures for specifying and/or estimating spatially structured correlation matrices.
Therefore, such implementations, and investigation of their small-sample performance, are required.

\subsection{Implications of our study}
Spatial auto-correlation is widespread in  evolutionary biology  data and it is likely that many studies based on the
partial Mantel tests  who concluded to the existence of an association between two sets of  variables
were based on an erroneously small p-value.
Spatial auto-correlation is perhaps the most common form of auto-correlation in evolutionary biology but any form of auto-correlation would lead to the same issue. In particular,  the case of two variables displaying some form of phylogenetic signal
is subject to the same issue.
The range of questions where using the partial Mantel test instead of the simple Mantel test alleviates the statistical
issues discussed here is very limited (presence of a deterministic linear trend and no other form of spatial structure).
Outside the situations enumerated in section {\em ``When are the Mantel tests valid statistical methods?''} above,
the Mantel tests are not validated statistical methods and any result based on them is dubious.
Mantel tests are widely used in the scientific community and while it has become increasingly obvious
that they are not well-grounded statistical methods, 
no computer
program implementing alternative methods have been developed.
This relates presumably to the several reasons: most users of the method have long been considering the partial Mantel test
as a ``safe'' alternative to the simple Mantel test,  the problem is not trivial,  useful methods should be able to deal with
all kinds of data (multivariate and categorical as well as quantitative data) and lastly,  'mainstream' spatial statistical methods
(which have known a tremendous development in the last twenty years,  most notably with the advent of hierarchical models
and  simulation methods) are traditionally more concerned with prediction than with testing.
Our study stresses the need for a general and well validated computer program.
As a final note we stress that the geostatistical modelling framework taken here is not the only meaningful framework
for the analysis of evolutionary biology data in this context. It is taken here mostly
because it allowed us to make statements
about statistical distributions {\em under some well defined assumptions},  which is often the missing
corner stone of analyses based on the Mantel tests.

\newpage

\section{ Funding}
This work has been supported by Agence Nationale pour la Recherche,  France,  under
project EMILE (grant ANR-09-BLAN-0145-01) and the Danish Centre for Scientific Computing (grant 2010-06-04).

\section{ Acknowledgements}
The first author is grateful to Murat Kulahci for constructive comments on an earlier draft and
 to the participants of the 9th French-Danish workshop in Spatial Statistics and Image Analysis in Biology in Avignon 
for stimulating comments.
\newpage
\bibliography{/media/SSD_Gilles/biogillesg/com/biblio/biblio,/media/SSD_Gilles/biogillesg/com/biblio/gilles}

\end{document}